\documentclass{article}
\usepackage{amsfonts}
\usepackage{amssymb}
\usepackage{amsmath}
\usepackage{caption}
\usepackage{graphicx}

\newif{\ifcomentarios}
\comentariosfalse

\begin{document}

\author{Pedro Ribeiro de Almeida, Vitor Hirata Sanches, Carla Goldman \\
Instituto de F\'{\i}sica - Universidade de S\~{a}o Paulo, CEP
05508-090, S\~{a}o Paulo-SP, Brasil.} 
\title{Balancing the Benefits of Vaccination: an \textit{Envy-Free}
Strategy. }
\date{June 2023 }
\maketitle

\begin{abstract}
The Covid-19 pandemic revealed the difficulties of vaccinating a population
under the circumstances marked by urgency and limited availability of doses
while balancing benefits associated with distinct guidelines satisfying
specific ethical criteria (J.W.Wu, S.D. John, E.Y. Adashi, Allocating
Vaccines in the Pandemic: The Ethical Dimension, The Am. J. of Medicine
V.33(11): 1241 - 1242 (2020)). We offer a vaccination strategy that may be
useful in this regard. It relies on the mathematical concept of
envy-freeness. We consider finding balance by allocating the resource among
individuals that seem to be heterogeneous concerning the direct and indirect
benefits of vaccination, depending on age. The proposed strategy adapts a
constructive approach in the literature based on Sperner%
\'{}%
s Lemma to point out an approximate division of doses guaranteeing that both
benefits are optimized each time a batch becomes available. Applications
using data about population age distributions from diverse countries suggest
that, among other features, this strategy maintains the desired balance
throughout the entire vaccination period.

\medskip

\textbf{Keywords:} pandemic preparedness, balanced vaccine allocation,
decision making process, envy-free division, direct and indirect benefits,
Sperner%
\'{}%
s Lemma, cake-cutting.
\end{abstract}

\section{\protect\bigskip Significance Statement}

Direct and indirect benefits of vaccination are related to decreasing the
severity of the individual%
\'{}%
s symptoms and to decreasing the spreading of the disease due to collective
effects. The share with which a single dose contributes to each type of
benefit may depend, among other conditions, on the age of the individual
that receives it. This imposes difficulties in optimizing allocation
guidelines that aim to support individual needs while controlling
transmissibility. We offer a strategy of vaccination that may balance these
two aspects based on the mathematical concept of envy-freeness. The present
study revealed the efficiency of such a strategy and its tendency to
equalize the benefits of vaccination locally, within a country, and among
countries presenting the most diverse age distribution profiles.

\section{Introduction}

The unprecedented situation in which a vaccine was successfully developed
amid the disease pandemic, the case of Covid-19, brought about urgent
questions related to the possible strategies for allocation of the doses
made available gradually in very small batches that do not cover the entire
population in a community all at once \cite{who protocols}, \cite{WHO}.
Given the high transmissibility of the virus and widespread infection, the
problems associated with vaccine allocation highlighted the urgent need to
elaborate and put in practice certain guidelines that best satisfy a given
set of ethical requirements \cite{ethical 1},\cite{science 1}. In many
countries, the prioritization followed protocols suggested by the WHO \cite%
{who protocols} to allocate the first doses that became available to the
oldest and to those with comorbidities since these individuals are the most
likely to develop severe forms of the disease \cite{clinical}, \cite%
{clinical 1}. Other groups at maximum risk such as health care workers and,
in some places, members of disadvantaged groups deprived of minimal
protection against direct exposure to the virus, the homeless for instance,
in addition to members of indigenous populations and isolated small
communities, among others, have also been eligible for doses of the vaccine
from the first batches \cite{who protocols}. \ 

It is worth noticing that, after the most elderly and other most-at-risk
groups receive their doses, virtually the entire population remains
unvaccinated while new doses continue to be available in very limited
numbers. In the remaining susceptible individuals, comprising the large
majority of the population, the ability to transmit the disease and the
severity of the symptoms are widely dispersed, and these generally correlate
with age. We restrict the contribution offered here to this scenario.

The elderly within this remaining population would still be mostly benefited
directly from receiving the doses because they tend to develop the most
severe forms of the disease compared to the younger that are likely to
present with only mild symptoms, although this is not a rule \cite{clinical}%
. On the other hand, due to their mobility and intense daily activity,
younger people have a major capacity to transmit the virus compared to that
of the elderly \cite{ethical 1}. Therefore, vaccinating younger people would
greatly benefit the entire population, as an indirect effect.

Direct and indirect effects of vaccination, regarding mainly the interplay
between decreasing disease severity and its transmissibility, raise
questions about the possibility of balancing these two factors that seem to
compete with each other in making decisions about allocating vaccine doses 
\cite{ethical 1}\textbf{.}\emph{\ }We believe that any solution to this
problem should comprise the following points: 1) a measure to evaluate
proximity to a balanced condition that enables comparing results among
different strategies of vaccination, and \ 2) a methodology to implement
dose allocation that maintains such balance on time until vaccine coverage
of the entire population is achieved.

We argue that this can be approached following an \textit{envy-free} type of
strategy for a fair division of doses among individuals possessing different
utilities. The concept of envy-freeness is often illustrated in the
literature by the classical cake-cutting problem \cite{how to cut a cake}, 
\cite{constructive}. This refers to a partition among agents of a certain
resource (the cake), generally heterogeneous, such that each one of the
agents evaluates their parts as being the best among the parts chosen by the
others. The heterogeneity of the resource is usually expressed through a
utility function that assigns to the different parts of the cake, different
values satisfying additivity. In general, each agent has its own utility
function. Here, we use the notion of utility to quantify both the direct and
indirect benefits of vaccinating the diverse age groups of the population.
We then formulate such a strategy to address the case of vaccine dose
allocation by the agents to the individuals adapting the constructive
approach presented in \cite{Su} and reviewed in \cite{Playful Introduction}
which is based on Sperner\'{}s Lemma. For this, we assume that the vaccine
is a desirable good and also that individuals and vaccine doses can be
conceived as divisible quantities being represented by densities, defined
appropriately. Accordingly, each age group receives a score named \textit{%
utility} in agreement with the various views and plans of certain
consultants or \textit{counselors} expressing their priority criteria in
line with the current public policies in the considered community.

We examine the case at issue regarding transmissibility and severity of the
disease as a prototype to explain our ideas, although the model is not
restricted to it. In keeping with this, it will be sufficient to consider
the expertise of only two counselors, each one in charge of scoring all
individuals according to their ages. One of the counselors referred to as $%
C_{A}$ (Ana), is an expert in predicting the ways of spreading the disease.
The other counselor, $C_{B}$ (Bob), is an expert in disease symptomatology.%
\emph{\ } Specifically, $C_{A}$ represents the allocation guideline that
accounts for the benefit of vaccinating to control transmission - the
indirect effect of vaccination. $C_{B}$ represents another allocation
guideline that accounts for the benefit at the individual level - the direct
effect of decreasing the severity of the symptoms. Both $C_{A}$ and $C_{B}$
are interested in balancing the two aspects; none of them wants to dispute
vaccine doses.\emph{\ }Therefore, whenever a new batch becomes available to
this community, the doses will be allocated in such a way that each
counselor agrees on the distinct groups of individuals to be vaccinated to
optimize separately the benefit envisaged by each one. We claim that this
characterizes an \textit{envy-free} division of the vaccine doses.

The way that this may be accomplished is our main proposal and will be
developed in the following Sections. We emphasize that unlike \textit{%
utilitarian models} \cite{utilitarian models 1}, \cite{utilitarian models 2}%
, our strategy is not based on a single score system for which the
priorities for doses allocation are evaluated in terms of the total score
received by each individual from the different counselors. Rather, we
conceive the model in such a way that each counselor optimizes the benefit
according to their particular view. Our approach differs also from the 
\textit{reserve system} strategies \cite{reserved systems} for which the
total vaccine supply from each batch is distributed according to
pre-assigned proportions to certain reserved categories. In our model, the
proportions of doses attributed to the management of each counselor are
dynamic quantities, resolved along the process.

Our proposal is presented in Section \ref{model}. Illustrative examples are
considered in Section \ref{results}. We compare the results from the
application of the \textit{envy-free }strategy\textit{\ }using data
comprised of certain population age distributions, with those predicted by
the other two strategies examined, referred to as \textit{oldest-first,} and 
\textit{maximize-benefit}, as detailed below. We have also considered a
strategy named \textit{minimize-benefit} to set a scale to measure the
efficiency with which the benefits are acquired by each strategy. These
comparative results indicate that the \textit{envy-free} leads to a
considerable improvement in keeping the benefits related to $C_{A\text{ \ }}$%
and $C_{B\text{ \ }}$close together over time conferring support to this
strategy as a way to pursue the desired balance. A discussion of these
results and considerations about the extent of the applicability of the
model are presented in Section \ref{discussion}. In the Supplementary
Material we outline the numerical procedure used to implement the model
dynamically.

\section{Model\label{model}}

Our formulation is decomposed into two parts. The first part consists of
preliminary definitions to build up the relevant simplex\ as a basis for the
choice of individuals to be vaccinated at each time. It is assembled using
accessible data about population age distribution, taken in connection with
the utility attributed to all groups of individuals by each counselor. The
second part consists in building up the dynamics that drives this choice to
achieve the required balance between the two guidelines.

\subsection{The simplex\label{simplex}}

The age group distribution in a community with $N$ susceptible individuals
will be considered for the account of two counselors $C_{A}$ and $C_{B},$
each one of them endowed with a utility\textit{\ function} $\rho _{\eta
}(I), $\emph{\ }$\eta =A,B$ hypothesized in such a way to attribute a score
to each individual according to the corresponding age-related priority
criteria. This can be performed by ordering these $N$ individuals in such a
way that their age $I(x)$ is a monotonic increasing surjective function of
their positions\emph{\ }$x\in \mathbb{Z} \cap \lbrack 0,N].$

In order to build up the \textit{1-simplex} of interest over which we
perform our considerations about the choices of the counselors, we map the
interval $[0,N]$ into the interval $[0,1]$ and variable $x$ into a real
variable $y=x/N$ such that $y\in \lbrack 0,1].$ The age at position $y$
within this map shall now be calculated as $I(yN).$ We may assume that all
individuals within the same age group are equally valued by each counselor
though most likely, the value\ varies between counselors. It will be
convenient though, to deal with continuous utility functions $\rho _{\eta
}(I(yN))\equiv \rho _{\eta }(y)$ represented by a combination of smoothed
step functions for each counselor $\eta =A,B$, as detailed in Section \ref%
{methods}. The \textit{utility densities} $u_{\eta }(y)$ defined for all $%
y\in \lbrack 0,1]$ as 
\begin{equation}
u_{\eta }(y)=\frac{\rho _{\eta }(y)}{\int_{0}^{1}\rho _{\eta }(y))dy}
\label{normalized ui}
\end{equation}%
are the functions that allow for considerations about \textit{envy-free}
divisions based on Sperner\'{}s Lemma, as will be explained next. Observe
also that any region $\omega $ of the considered simplex may be decomposed
into a number $M$ of disjoint sub-regions $\{v_{j}\},$ $j=1,2,...M$. Each $%
v_{j},$ extended between endpoints $y_{jI}$ (initial) and $y_{jF}$ (final)
with $y_{jI}<y_{jF}$, comprises a number $\left[ \left( y_{jF}-y_{jI}\right)
N\right] $ of individuals, where the notation $\left[ z\right] $ indicates
the integer part of the real number $z$.

\bigskip

\subsection{The dynamic\label{dynamic}}

Consider a population that at a certain instant of time $t$ comprises $N(t)$
susceptible individuals to whom a batch of $V<N$ vaccine doses shall be
allocated. We suppose that the availability of the batches occurs at a
certain frequency of $1/T$ until the entire population is vaccinated. We
also assume that individuals achieve full protection after receiving a
single dose. The time $t$ shall then be better measured in terms of the
interval $T$ between batches as $t=nT,$ for $n\in \mathbb{Z}_{+}.$ The
question posed here regards the choice of the $V$ individuals to receive the
doses at each time $t$ in order to balance the current guidelines.

We think of two different priority criteria suggested by two \textit{%
counselors} $C_{A}$ and $C_{B}$\ expressing different opinions about how one
should rank the population in the community to guide this choice. The 
\textit{utility density functions} $u_{A}(y)$\ and $u_{B}(y),$ $y\in \lbrack
0,1]$\ conceived, respectively, by $C_{A}$\ and $C_{B}$\ assume nonnegative
real values and represent a measure of the relevance for vaccinating the
individuals ordered according to some rule. To present the methodology, we
choose to order the individuals by their age, although this does not exclude
any other possibility. Such an ordered list of individuals mapped into the
interval $[0,1]$\ defines the \textit{1- simplex} as detailed above. We aim
to present a fair division strategy of an \textit{envy-free} class\textbf{\ }%
through which the choice of the $V$ individuals at each time balances the
perspectives of the two counselors in the best possible way.\emph{\ }The
proposal offers an approximate solution extending the constructive approach 
\cite{constructive} based on Sperner\'{}s Lemma as presented in \cite{Su}
and reviewed in \cite{Playful Introduction}.

From the utility density functions $u_{\eta }(y)$, $\eta \in \{A,B\}$ we
define the \textit{benefit }$\mathcal{U}_{\eta }^{\omega (t)}$\textit{\ }
according to the referred counselor perspective, that results from
vaccinating the individuals inside a region $\omega (t)$ of the simplex at
time $t$: 
\begin{equation}
\mathcal{U}_{\eta }^{\omega (t)}\equiv \int_{\omega (t)}u_{\eta }(y)dy\text{ 
}  \label{Ui}
\end{equation}%
The total benefit that is reached after vaccinating the entire population of
susceptible amounts to one, according to both counselors:%
\begin{equation}
\mathcal{U}_{\eta }=\int_{0}^{1}u_{\eta }(y)dy=1.  \label{normalize}
\end{equation}

We proceed\ by partitioning the 1-simplex, at a time $t,$ into a number $d$
of identical parts, each of which comprised between a pair of neighbor
points $(p_{i},p_{i+1})$ at the positions $(y_{p_{i}},y_{p_{i+1}})$,
respectively, with $y_{p_{j}}=j/d,$ for $j=\{0,1,2,...d\}$. We then assign
to the endpoint $p_{0\text{ \ }}$at $y_{p_{0\text{ \ }}}=0$\ a label
arbitrarily chosen between $A$\ and $B$\ so called in reference to the
counselors, and then proceed by labeling each of the following points $p_{i%
\text{ \ }}$of the sequence as $A$\ or $B$, alternately.

Observe that each point $p_{i}$ at $y_{p_{i\text{ \ }}}$splits the ordered
population into two parts, part $I$ on the left of $p_{i}$ and part $II$ on
the right of $\ p_{i}$, comprising respectively $N_{I}^{i}$ and $N_{II}^{i}$
individuals at time $t$, such that

\begin{equation}
\begin{split}
N_{I}^{i}& =\left[ y_{p_{i}}N\right] \\
N_{II}^{i}& =\left[ (1-y_{p_{i}})N\right]
\end{split}
\label{p (1-p)}
\end{equation}

At each of these points $p_{i}$ we also consider splitting the batch of
vaccines available at a certain time into two parts, $V_{I}^{i}$\ and $%
V_{II}^{i}$\ . We choose $V_{I}^{i}$\ and $V_{II}^{i}$\ \ proportional to $%
N_{I}^{i}$\ and $N_{II}^{i},$ respectively:

\begin{equation}
\begin{array}{l}
V_{I}^{i}=\left[ y_{p_{i}}V\right] \\ 
V_{II}^{i}=\left[ (1-y_{p_{i}})V\right]%
\end{array}
\label{V1V2}
\end{equation}

To the extent that the simplex is arranged in this way, both quantities,
i.e. individuals and vaccine doses, are evaluated using the single
continuous variable $y$. This allows each counselor to express, at the
corresponding labeled points $p_{i}$, what would be their preferential side
to proceed with vaccination. We assume that $V_{I}^{i}$ and $V_{II}^{i}$\
are intended necessarily to vaccinate individuals, respectively, on sides $I$
and $II$ defined for each $i$. Explicitly, to maximize benefit at each $A$%
-labeled point $p_{i}$, counselor $C_{A}$ is asked to express her
preference, based on $u_{A}$, about vaccinating $V_{I}^{i}$ individuals on
the side $I$ or $V_{II}^{i}$ individuals on the side $II.$ The same for
counselor $C_{B}$ at each $B$-labeled point $p_{i},$based on $u_{B}$. One
should notice that it is implicit in this procedure, regardless of the
counselors\'{} choices, that neither of them would be able to vaccinate the
entire population at once with the corresponding amount $V_{I}^{i}$ or $%
V_{II}^{i}$ of doses available on each side. Moreover, at an $A$\emph{-}%
labeled point$\ p_{i}$ where counselor $C_{A}$\ is in charge of choosing the
side and decides for say, side $II,$ she is supposed to make use of all of
the $V_{II}^{i}$ doses pre-assigned to that side. This implies that
counselor $C_{B}$ would necessarily vaccinate individuals on the other side
using all of the $V_{I}^{i}$ doses, even though it may not be his
preferential\'{}s. Despite this, $C_{B}$\ would look for individuals
amounting to $V_{I}^{i}$ that are best to be vaccinated on the side $I$\
according to his utility density function. The same will be followed by
counselor $C_{A}$\ after $C_{B}$\ has expressed his preferential choices at
each $B$-labeled point $p_{i}$.

Accordingly, the counselor in charge at each point $p_{i}$, regardless of
being labeled $A$ or $B$, ends up vaccinating exclusively at one of the
sides. Nonetheless, they would benefit from vaccination on both sides. Since
both utility density functions assume nonzero values along the entire
simplex, each counselor must account for a \textit{benefit coupled} with the
other\'{}s choice. In the example above we understand that, at that
particular point $p_{i},$ counselor\emph{\ }$C_{A}$\emph{\ }has chosen side%
\emph{\ }$II$\emph{\ }and the best sub-region to vaccinate the $V_{II}^{i}$
individuals at that side. This choice is foreseen after evaluating the total
benefit from $u_{A}$ composed of: (i) the amount obtained from $u_{A}$\ at a
sub-region of $II$ comprising $V_{II}^{i}$\ individuals that\ have been
chosen according to her utility function $u_{A}$, and (ii) the \textit{%
coupled benefit} that corresponds to the amount obtained from $u_{A}$ at a
sub-region of $I$ comprising $V_{I}^{i}$\ individuals that have been chosen
by suggestion of $C_{B},$\ based on his utility density $u_{B}.$ She
concluded that the sum of (i) and (ii) is greater than the amount she would
have obtained if she has chosen to vaccinate on the side $I$\ and received
the coupled benefit from side $II.$ For this, one must assume that both
counselors know each other\'{}s utility density functions.

To extend $(i)$ and $(ii)$ to arbitrary choices, it will be useful to define
for $\Gamma \in \{I,II\}$\text{ and }$\eta \in \{A,B\},$ the interval $%
\Omega _{\eta }^{\Gamma }(p_{i},t),$ as the sub-region on the side $\Gamma $
of the simplex with respect to point $p_{i}$ where counselor $\eta $
evaluates the maximum benefit from $u_{\eta }$ at time $t$. According to
this, counselor $C_{A}$ looks for the largest between the \textit{total} 
\textit{benefit} $U_{A}^{I}(p_{i})$ and $U_{A}^{II}(p_{i})$ that, recalling (%
\ref{Ui}), are defined as:

\begin{equation}
\begin{array}{c}
U_{A}^{I}(p_{i},t)\equiv \mathcal{U}_{A}^{\Omega _{A}^{I}}+\mathcal{U}%
_{A}^{\Omega _{B}^{II}}%
\end{array}
\label{total utility A em I}
\end{equation}%
and

\begin{equation}
\begin{array}{c}
U_{A}^{II}(p_{i},t)\equiv \mathcal{U}_{A}^{\Omega _{A}^{II}}+\mathcal{U}%
_{A}^{\Omega _{B}^{I}}%
\end{array}
\label{total utility A em II}
\end{equation}%
If $U_{A}^{I}(p_{i},t)\geqslant U_{A}^{II}(p_{i},t)$ she decides for side $I$%
, otherwise she decides for side $II.$

Likewise, to decide which side to vaccinate at each $B$-labeled point$\
p_{i} $, according to his utility function, counselor $C_{B}$\ looks for the
largest between the \textit{total benefit} $U_{B}^{I}(p_{i},t)$ and $%
U_{B}^{II}(p_{i},t),$\ defined as:

\emph{\bigskip }%
\begin{equation}
\begin{array}{c}
U_{B}^{I}(p_{i},t)\equiv \mathcal{U}_{B}^{\Omega _{B}^{I}}+\mathcal{U}%
_{B}^{\Omega _{A}^{II}}%
\end{array}
\label{total utility B em I}
\end{equation}%
and%
\begin{equation}
\begin{array}{c}
U_{B}^{II}(p_{i},t)\equiv \mathcal{U}_{B}^{\Omega _{B}^{II}}+\mathcal{U}%
_{B}^{\Omega _{A}^{I}}%
\end{array}
\label{total utility B em II}
\end{equation}%
If $U_{B}^{II}(p_{i},t)\geqslant U_{B}^{I}(p_{i},t)$ he decides for the side 
$II$, otherwise he decides for the side $I.$

The example discussed above corresponds to the case for which the
pre-evaluation of the benefit by $C_{A},$ at the considered point $p_{i},$
resulted in $U_{A}^{II}(p_{i},t)>U_{A}^{I}(p_{i},t).$

We finally observe that even though side $I$ has no individuals to be
vaccinated at the end-point $p_{0}$ at $y_{p_{0}}=0,$ the counselor in
charge there might have two options: either to let the other vaccinate on
side $II$ using the entire amount $V$ of doses, or to vaccinate the $V$
individuals on side $II$. For example, if \ the point $p_{0}$ is $A$-labeled
then counselor $C_{A}$ will still be in charge to decide about her
preferential side based on the largest between

\text{ \hspace{0.51in}}

\begin{equation}
\begin{split}
U_{A}^{I}(p_{0},t)& =\mathcal{U}_{A}^{\Omega _{B}^{II}(p_{0},t)}\hspace{%
0.51in} \\
& \text{and} \\
U_{A}^{II}(p_{0},t)& =\mathcal{U}_{A}^{\Omega _{A}^{II}(p_{0},t)}\text{%
\hspace{0.51in}}
\end{split}
\label{UA p=0}
\end{equation}%
On the contrary, if the end-point $p_{0}$ is $B$-labeled then counselor $%
C_{B}$ would select the side based on the largest between

\begin{equation}
\begin{split}
U_{B}^{I}(p_{0},t)& =\mathcal{U}_{B}^{\Omega _{A}^{II}(p_{0},t)}\text{%
\hspace{0.51in}} \\
& \text{and} \\
U_{B}^{II}(p_{0},t)& =\mathcal{U}_{B}^{\Omega _{B}^{II}(p_{0},t)}\text{%
\hspace{0.51in}}
\end{split}
\label{UB p=0}
\end{equation}

Since at $p_{0}$ the values reached by $u_{A}$ at $\Omega _{A}^{II}(p_{0},t)$
are higher than or at least equal to the values reached by $u_{A}$ at $%
\Omega _{B}^{II}(p_{0},t)$ then $U_{A}^{II}\geqq U_{A}^{I}$. Similarly,
since the values reached by $u_{B}$ at $\Omega _{B}^{II}(p_{0},t)$ are
higher than or at least equal to the values reached by $u_{B}$ at $\Omega
_{A}^{II}(p_{0},t)$ then $U_{B}^{II}\geqq U_{B}^{I}$. Therefore, the
counselor in charge at $p_{0}$ will oneself prefer to indicate the
individuals to be vaccinated, and this would happen on side $II$, instead of
leaving vaccination up to the other counselor. Using similar arguments, one
finds that for $p_{1}$ at $y_{p_{1}}=1$ either one of the counselors would
choose side $I$. Therefore, whoever counselor at $p_{0},$ would choose the
right side whereas whoever counselor at $p_{1},$would choose the left side.
These conclusions assure that the conditions under which Sperner's Lemma
holds are fully satisfied by the simplex defined above.

Finally, after the two counselors have expressed their preferential sides at
each of the corresponding points $p_{i}$ and the simplex looks like that
sketched in \textbf{Figure 1} it allows one, through simple visual
inspection, to list all pairs of consecutive points, referred here
generically as $(p_{L},p_{R}),$ such that the counselor at the point on the
left $p_{L}$ has expressed a preference to vaccinate on one of the sides say
on side $II,$ while the counselor at the point on the right $p_{R}$ has
expressed a preference to vaccinate on the opposite side, i.e. side $I$. 

The existence of at least one such pair of points is ensured by Sperner's
Lemma. Accordingly, for sufficient large partition\emph{\ }$d$, an internal
point $p^{\ast }$ of the interval defined by any of these pairs will
approximate a position at which the preferred sides of the two counselors
are opposite to one another. The choice of any of those points $p^{\ast }$
(if more than one) identifying opposite preferred sides for each counselor
to allocate the available vaccine doses, characterizes an approximate 
\textit{envy-free} division for which either

\begin{equation}
U_{A}^{I}(p^{\ast })\geqslant U_{A}^{II}(p^{\ast })\text{ and }%
U_{B}^{II}(p^{\ast })\geqslant U_{B}^{I}(p^{\ast })
\label{condicao EF maior}
\end{equation}%
or

\begin{equation}
U_{A}^{I}(p^{\ast })\leqslant U_{A}^{II}(p^{\ast })\text{ and }%
U_{B}^{II}(p^{\ast })\leqslant U_{B}^{I}(p^{\ast })
\label{condicao EF menor}
\end{equation}

In order to carry on this strategy until all susceptible individuals in the
population have the opportunity to get their doses, it is assumed that the
procedure described above is repeated at each time $t$ when a new batch
containing $V$\ doses becomes available. For simplicity, we consider the
unrealistic case for which $V$\ does not change along the entire process. On
each of these occasions, the simplex must be re-scaled and the utility
densities attributed accordingly to the set of individuals mapped again into
the interval $[0,1],$\ after excluding those already vaccinated with the
doses from the previous batch.

We present a numerical study using this procedure for analyzing the time
evolution of the benefit in selected population age distributions. The
utility functions are written using an arbitrary scale to mimic counselors`
general guidance. \ The results are compared with those produced by
strategies specified in the following as \textit{maximize-benefit}, \textit{%
oldest-first} in addition to a \textit{minimize-benefit} strategy introduced
to set a scale for efficiency. The \textit{maximize-benefit} strategy looks
for distributing the doses to the groups of individuals for which the sum of
the two utilities is maximized. The \textit{minimize-benefit} strategy does
the opposite. Under the \textit{oldest-first} strategy, the doses available
are fully distributed to the oldest individuals present at the time,
approaching the current procedure adopted by many public health systems. Our
findings are shown in the next Section.

\section{Results\label{results}}

The time evolution of benefits acquired by applying each of the three
strategies mentioned above is studied through numerical simulations. The
methodology outlined (Supplementary Material) has been developed
specifically to accomplish this. We use data for population age distribution
of the countries indicated in \cite{demographic}. For comparing the outcomes
from the diverse strategies, the population of each country is divided into
four age groups $I_{k},$ $k=1,2,3,4$, comprising, respectively, individuals
from 0 to 14 years old $(I_{1})$, from 15 to 24 years old $(I_{2})$, from 25
to 64 years old $(I_{3})$, and those that are 65 years old or above $(I_{4})$%
. Any other division could have been considered. Before proceeding into the
normalization, each counselor $\eta $ assigns to each of these groups a
utility value according to their particular priority criteria. Our choices
are conceived using an arbitrary scale to set the quantities employed in the
examples. These are indicated by the components of the vectors $\Psi
_{(A)}^{(1)}=\left( 12,16,4,1\right) $ or \ $\Psi _{(A)}^{(2)}=\left(
7,23,2,1\right) $ \ for $C_{A},$\ and $\Psi _{(B)}^{(1)}=$\ $\left(
1,4,12,16\right) $ or $\Psi _{(B)}^{(2)}=\left( 1,2,7,23\right) $ for $C_{B}$%
, which assume non-zero positive values and are independent of the number of
individuals in each age group, characteristic of each community. These are
then applied on Eq.(\ref{normalized ui}) to build the utility density
functions $u_{\eta }(y)$\ for the diverse age distributions. We examine the
four combinations: $\Psi _{(A)}^{(1)}$\ and $\Psi _{(B)}^{(1)}$\ (Default), $%
\Psi _{(A)}^{(1)}$\ and $\Psi _{(B)}^{(2)}$\ (Symptomatology.), $\Psi
_{(A)}^{(2)}$\ and $\Psi _{(B)}^{(1)}$\ (Transmissibility), $\Psi
_{(A)}^{(2)}$\ and $\Psi _{(B)}^{(2)}$\ (Concentrated). The idea is to test
the choices of the counselors as the utilities become concentrated on the
groups that each one of them finds the most priority to compare with the
cases for which the utilities are less concentrated in a single group
(Default). The utility density functions obtained using the population age
distribution of the U.S. are shown in \textbf{Figure 2}. Analogous results
have been obtained for the remaining countries (not shown). We emphasize
that the assignments above for each $\Psi _{(\eta )}^{(1,2)}$ represent
possible choices to compare the achievements of the different strategies and
combinations of utilities considering the diverse age distributions. Any
other possibility would be feasible, depending on the interests and
attributions of the counselors.

Given $u_{\eta }(y)$ for each country and for each counselor, the simulated
dynamic compares the outcomes using three different strategies to allocate
doses namely, \textit{Envy-Free}, \textit{Oldest-First,} and \textit{%
Maximize-Benefit} one at the time, until vaccination ends. \textit{\ }

Under \textit{envy-free}, the preferred side\text{ }of\text{ each of the two
counselors at }$p^{\ast }(t)$\text{ are opposite to one another}. Yet,
because for a given $\eta ,$ $U_{\eta }^{\Gamma }(p^{\ast })$ accounts also
for the coupled benefits associated with the other\'{}s choice [\ref{total
utility A em I} - \ref{total utility B em II}], the total region $\Omega
(t)_{\text{\textit{EF}}}$ to be vaccinated at each time $t$ is necessarily
composed of two or more disjoint segments spanned on both sides, the same
for $C_{A}$ and for $C_{B}.$ That is, $\Omega (t)_{\text{\textit{EF}}%
}=\Omega _{A}^{I}\cup \Omega _{B}^{II}$ if $C_{A}$ preferred side $I$ and $%
C_{B}$ preferred side $II,$ or $\Omega (t)_{\text{\textit{EF}}}=\Omega
_{A}^{II}\cup \Omega _{B}^{I}$ if $C_{A}$ preferred side $II$ and $C_{B}$
preferred side $I.$

Under \textit{oldest-first}, the focus is on the protection of the elderly.
In this case, the choice of the fraction of individuals to be vaccinated
with available doses is based exclusively on the distribution of the age
groups. The preference is always for the $V$ most elderly which fraction $%
v(t)=V/N(t)$ comprises a one segment region $\Omega (t)_{\text{oldest}}$ of
the simplex at each time $t$. The \textit{maximize-benefit} strategy, on the
other hand, is based on the choice of the fraction $v(t)$ of individuals
comprising a region $\Omega (t)_{\max }$, eventually composed of disjoint
regions, where the benefit achieved by adding the utilities from the two
counselors is maximized.

Using definition (\ref{Ui}), we express the average increment to the benefit
achieved at time $t$ as: 
\begin{equation}
\mathcal{U}_{\text{\ \ }}^{\Omega (t)}=\frac{1}{2}(\mathcal{U}_{A}^{\Omega
(t)}+\mathcal{U}_{B}^{\Omega (t)})  \label{U(t)}
\end{equation}%
for all strategies, such that\emph{\ }$\Omega (t)\in \left\{ \Omega (t)_{%
\text{\textit{EF}}}\ ,\Omega (t)_{\text{oldest}}\ ,\Omega (t)_{\max },\Omega
(t)_{\text{rand}}\right\} $\emph{.} These include a random vaccination
process considered for comparison purposes, under which the fraction $v(t)$
of doses is offered to a randomly chosen fraction $\Omega (t)_{\text{rand}}$
of the simplex.

In all cases, the simulations run for an initial population comprising $%
N(0)=10^{4}$ individuals and fixed vaccine batches of $V=10^{2}$ doses each.%
\emph{\ }The simplex built at every iteration time to follow the envy-free
strategy was partitioned using $d=100$\ that ensures convergence of the
results, as suggested by the study depicted in \textbf{Figure S2}. The whole
procedure intends to find the regions $\Omega (t)$ to distribute the doses
at each iteration time, that conform with each of the considered strategies. 
\textbf{Figures (3-7)}\emph{\ }show results considering utilities combined
as $\tau _{(A)}^{(1)}$\ and $\tau _{(B)}^{(1)}$\ (Default). The other
combinations are also examined and the results are collected in \textbf{%
Figures (8-9)}. The time behavior of the increments $\mathcal{U}_{A}^{\Omega
(t)},\mathcal{U}_{B}^{\Omega (t)}$\ and $\mathcal{U}_{\text{\ \ }}^{\Omega
(t)}$ in\textbf{\ Figure 3}\ are for the population of the U.S. All other
distributions that we have examined exhibited the same patterns (results not
shown). \textbf{Figure 4}\ exhibits the results for selected countries, as
listed. Each point represents the time average of the differences (absolute
values) $\overline{\Delta \mathcal{U}\text{ }}$ between the increments to
the benefit achieved by each of the two counselors, 
\begin{equation}
\overline{\Delta \mathcal{U}}\equiv \frac{1}{\tau }\displaystyle%
\sum_{t=0}^{\tau }\left\vert \mathcal{U}_{A}^{\Omega (t)}-\mathcal{U}%
_{B}^{\Omega (t)}\right\vert  \label{time_average _difference}
\end{equation}%
evaluated over the time interval $\tau $\ encompassing the entire
vaccination period, for the diverse strategies.

\textbf{Figure 5} illustrates with the example of the U.S., the results
obtained for the time evolution of the cumulative benefits $\Phi _{\eta }(t)$%
: 
\begin{equation}
\Phi _{\eta }(t)=\displaystyle\sum\limits_{t{\acute{}} =0}^{t}\mathcal{U}%
_{\eta }^{\Omega (t{\acute{}})}  \label{cumulative benefit}
\end{equation}%
for each of the two counselors $\eta =A,B$, and the mean:%
\begin{equation}
\Phi (t)=\frac{1}{2}(\Phi _{A}(t)+\Phi _{B}(t))  \label{average_cum}
\end{equation}%
The outcomes obtained for all selected countries exhibited a similar pattern
(results not shown).

\textbf{Figure 6} depicts the time average of the differences (absolute
values) between the contributions to the cumulative benefits of the two
counselors, evaluated for all the countries listed:%
\begin{equation}
\overline{\Delta \Phi }=\frac{1}{\tau }\displaystyle\sum_{t}\left\vert \Phi
_{A}(t)-\Phi _{B}(t)\right\vert  \label{time average diff cum}
\end{equation}

\textbf{Figure 7} shows the corresponding results for time average $%
\overline{\Phi }$\ of $\Phi (t)$ (\ref{average_cum}):\emph{\ } 
\begin{equation}
\overline{\Phi }=\frac{1}{\tau }\displaystyle\sum_{t}\Phi (t)
\label{time_average_cum}
\end{equation}%
These include the results for the minimize-benefit which is worth
considering here precisely because it offers a lower bound to set a scale
that allows one comparing outcomes, as we discuss next. \textbf{Figure 8}
merges the results for the averages of cumulative benefits $\overline{\Delta
\Phi }$ and $\overline{\Phi }$ using the considered strategies and
combinations of utilities listed above extended for 236 countries (not
specified).

\section{Discussion and concluding remarks\label{discussion}}

The realistic case addressed here is that of deciding about strategies for
allocating vaccine doses that become available to a community at a certain
frequency but in very limited quantities. In the example used, we consider
two guidelines to drive allocation. The first focuses on the direct benefit
of decreasing the severity of the symptoms. The second focuses on the
indirect benefit of decreasing transmission. We approach this situation by
representing each of these guidelines as the priority of a qualified
counselor in scoring the entire population ordered by age. Assuming that
full protection of an individual is achieved after a single dose, the
challenge is to select the group of individuals to be vaccinated every time
a new batch becomes available to balance these two contributions. The
difficulty relies on the fact that, in general, the amount by which a given
vaccinated individual contributes to each of the benefits differ from each
other. On the contrary, if both benefits were of the same magnitude, any
strategy would result in a balanced condition. We claim that the strategy
based on an \textit{envy-free} division for dose allocation, as outlined
above, offers a suitable and efficient choice to achieve such a balance in
unpaired cases, as exemplified by the considered utilities. Our approach
adapts the constructive analysis of the classic cake-cutting division
problem \cite{Su} to conceive distributing doses optimizing the benefits
envisaged by each counselor, which include the benefits coupled with the
other\'{}s choice.

Consistent with this, the results in \textbf{Figure 3} of a case study
certify that under the \textit{envy-free} strategy, the increments to the
benefit acquired at each time by each counselor remain very close together
until vaccination is completed. Such results contrast with those obtained
through \textit{oldest-first} and \textit{maximize-benefit} for which the
increments to the two benefits differ considerably from each other across
time. By adopting any of these two strategies, the selected regions of the
simplex for doses allocation, either $\Omega =\Omega _{\text{oldest}}$\ or $%
\Omega =\Omega _{\max }$\ along which $u_{A}(y)$\ and $u_{B}(y)$\ may assume
very different values, leads to unbalanced $U_{A}^{\Omega (t)}$ and $%
U_{B}^{\Omega (t)}$. This is also the case with the random procedure.\textbf{%
\ Figure 4} suggests a measure $\overline{\Delta \mathcal{U}}$\emph{\ }(\ref%
{time_average _difference}) for this imbalance averaged over time. The
results for the diverse strategies are depicted for each of the selected
countries. It shows that $\overline{\Delta \mathcal{U}}$ approaches null
values through the \textit{envy-free} strategy. Relatively large values are
obtained by applying the other two strategies and also by choosing the
regions at random.

Although each of the benefits accumulates to the unity at the same time, the
way that this is accomplished and the effects on the achievements of the two
counselors can differ considerably. In this respect, the comparative results
in \textbf{Figure 5} offer information about the efficiency with which the
benefits evolve under different strategies. This can be better seen by
interpreting cumulative data as the positions in time of the two particles
in the space of benefits driven, each of them, by the corresponding
counselor. Extending the analysis for the population age distributions of
the selected countries, as shown, \textbf{Figure 6} depicts the average
distance kept between these two particles in each case, until reaching their
common final position simultaneously. Large values indicate that on average,
one of the particles reached positions close to the goal considerably faster
than the other. That is, for such strategies, the two benefits evolve out of
sync over a considerably large period. This is the case for\textit{\
maximize-benefit} and \textit{oldest-first} in these examples. On the
contrary, the positions of the two particles under the \textit{envy-free}
remain very close together at each instant through the entire time interval,
suggesting that in addition to offering a way to promote a balance between
acquired benefits, the strategy offers also a way to balance the
instantaneous rates at which this happens. This is important for practical
purposes since the intervals between consecutive batches may be very large,
especially during the initial vaccination. The effects of a time delay
between the achievements of each of the two benefits may have devastating
consequences for the community. The random choice procedure offers balanced
rates, on average, although the instant rates differ considerably since the
sizes of the increments to the benefits alternate unbalanced.

Keeping with this kinematic interpretation, data in \textbf{Figure 7} refer
to the time averages $\overline{\Phi }$\emph{\ }of the positions of the
center of mass of the two particles achieved through the diverse strategies,
and for all of the selected countries. The \textit{maximize-benefit}
presents the largest average value, as expected. Even though the partial
benefits, i.e. the ones envisaged by each counselor, evolve at different
rates in this case, both of them reach large values within relatively short
times. \textit{Envy-free} is also efficient in accumulating benefits almost
as fast as the \textit{maximize-benefit} does. Apparently, in all cases, the 
\textit{oldest-first} and the random choice promote the worst results among
all the considered procedures, except for a strategy introduced here named 
\textit{minimize-benefit}. Under this, one looks for regions of the simplex
that minimize the total benefit at each time. Although very implausible to
be adopted in practice, this strategy is useful to consider in the present
analysis since it provides a lower bound to compare the efficiency of the
diverse strategies investigated. Accordingly, the average benefit
accumulated upon \textit{envy-free} is much closer to the quantity
accumulated by the \textit{maximize-benefit} than that accumulated by the%
\textit{\ minimize-benefit }(\textbf{Figure 7}). Random choice accumulates
benefits at an average rate between the \textit{maximize} and \textit{%
minimize}-\textit{benefit} strategies. The \textit{oldest-first} spreads its
contributions along the interval showing a strong dependence on the
population age distribution.

\textbf{Figure 8 }depicts the results for (a) the time averages of the
instantaneous difference $\overline{\Delta \mathcal{U}}$ and (b) the time
averages of the cumulative difference $\overline{\text{ }\Delta \Phi }$,
both against the average cumulative benefit\emph{\ }$\overline{\Phi }$. Each
point from a total of 236 composing a colored set, corresponds to the
population distribution of a given country (not identified). The emphasis is
given to the different combinations of utilities and strategies employed. In
all cases, the results are in line with the behavior depicted in \textbf{%
Figures 4, 6,} and \textbf{7}, for the utility pair named Default. The 
\textit{envy-free} strategy is unique in achieving the smallest differences $%
\overline{\Delta \mathcal{U}}$ and $\overline{\text{ }\Delta \Phi }$ among
all strategies and in producing total benefit at a rate that, on average, is
the closest to that achieved by \textit{maximize-benefit}. Although the 
\textit{maximize-benefit} (and in parallel, the \textit{minimize-benefit})
approaches the results for the \textit{envy-free} regarding the cumulative
difference $\overline{\text{ }\Delta \Phi }$, the dispersion of data
increases considerably in these cases. A surprising outcome from the study
in Figure 8 is that the \textit{envy-free} reveals a tendency to minimize
the dispersion of the distributions for both $\overline{\text{ }\Delta \Phi }
$\emph{\ }and\emph{\ }$\overline{\Phi }$ when compared to the corresponding
results achieved by the other strategies. For all pairs of utilities chosen,
the remaining strategies produced large dispersion either for $\overline{%
\text{ }\Delta \Phi }$\emph{\ }or\emph{\ }$\overline{\Phi }$, or for both.
Collectively, these results indicate that among all of the considered
strategies the \textit{envy-free} promotes a good balance between the
benefits envisaged by the two counselors over time, and also that this
happens at similar and relatively high rates at initial times resulting in
fast accumulation of benefits. In addition, it is the strategy that tends to
equalize the benefits of vaccination among diverse countries, which is
desirable within a scenario of a pandemic. We thus believe that the proposed
strategy fulfills the requirements stated in the introduction since it
maintains the balance in agreement to different measures comprising a) the
amount of benefit acquired at each time by each counselor, b) the efficiency
of the process given the speed with which the cumulative benefit approaches
limiting values, and c) the tendency to equalize the effects achieved by
distinct population age distributions.

We have assumed throughout that the only mechanism by which individuals are
removed from the simplex is through vaccination. We do not account for
varying vaccine efficacy or deaths, whether caused by the disease or by any
other reason across the vaccination period. Once the two counselors provide
the utilities, we predict the fraction of individuals from each age group,
and at each time, that should be vaccinated to guarantee the balance. The
results in \textbf{Figure 9} exemplify in the case of the U.S. the kind of
outcome provided by each of the four combinations of utilities, as
indicated. In all cases, individuals of $65+$ and those comprised within $%
15-24$ years old are indicated to be prioritized across the initial
batches.\ \ 

The effects of vaccine efficacy have been considered in previous studies
using optimization algorithms \cite{science3}, \cite{pnas21} in connection
to the evolution of age-stratified population models. In particular, an SEIR
(susceptible, exposed, infectious, recovered) model dynamics has been
considered for analyzing different scenarios for the choice of the age
groups at the initial period of vaccination \cite{science2}. Given the
proposal developed here, it might be interesting to conceive a vaccination
plan based on an interplay between the dynamic of the envy-free and that of
the SEIR model. Such a protocol would be able to minimize morbidity while
balancing benefits. \ \ \ \ \ \ \ \ \ \ \ \ \ \ \ \ \ \ \ \ \ \ \ \ \ \ \ \
\ \ \ \ \ \ \ \ \ \ \ \ \ \ \ \ \ \ \ \ \ \ \ \ \ \ \ \ \ \ \ \ \ \ \ \ \ \
\ \ \ \ \ \ \ \ \ \ \ \ \ \ \ \ \ \ \ \ \ \ \ \ \ \ \ \ \ \ \ \ \ \ \ \ \ \
\ \ \ \ \ \ \ \ \ \ \ \ \ \ \ \ \ \ \ \ \ \ \ \ \ \ \ \ \ \ \ \ \ \ \ \ \ \
\ \ \ \ \ \ \ \ \ \ \ \ \ \ \ \ \ \ \ \ \ \ \ \ \ \ \ \ \ \ \ \ \ \ \ \ \ \
\ \ \ \ \ \ \ \ \ \ \ \ \ \ \ \ \ \ \ \ \ \ \ \ \ \ \ \ \ \ \ \ \ \ \ \ \ \
\ \ \ \ \ \ \ \ \ \ \ \ \ \ \ \ \ \ \ \ \ \ \ \ \ \ \ \ \ \ \ \ \ \ \ \ \ \
\ \ \ \ \ \ \ \ \ \ \ \ \ \ \ \ \ \ \ \ \ \ \ \ \ \ \ \ \ \ \ \ \ \ \ \ \ \
\ \ \ \ \ \ \ \ \ \ \ \ \ \ \ \ \ \ \ \ \ \ \ \ \ \ \ \ \ \ \ \ \ \ \ \ \ \
\ \ \ \ \ \ \ \ \ \ \ \ \ \ \ \ \ \ \ \ \ \ \ \ \ \ \ \ \ \ \ \ \ \ \ \ \ \
\ \ \ \ \ \ \ \ \ \ \ \ \ \ \ \ \ \ \ \ \ \ \ \ \ \ \ \ \ \ \ \ \ \ \ \ \ \
\ \ \ \ \ \ \ \ 

Finally, it should emphasize that the model offered here is not in any
possible way restricted to the specific guidelines addressed above. These
have been selected as references to explain and illustrate the practice of
the method. Any other guideline could have been chosen to drive the
allocation of available units.\ In addition, because Sperner's Lemma can be
extended to more dimensions \cite{Su}, \cite{Playful Introduction}, this
opens the possibility to extend the constructive strategy described above to
approach more realistic situations in which there are more than two
guidelines defining priorities \cite{ethical 1}. We believe that this offers
an attractive perspective to resolve such complex problems, with the help of
careful and skilled counselors.

\bigskip \newpage

{\Huge Supplementary Material}

\bigskip

{\LARGE Methods} \label{methods}

\bigskip

Here, we sketch the algorithm we have developed to find the envy-free
division for vaccine allocation, given a pair of utility density functions.

The time $t=nT$ of the $n^{th}$ iteration is measured in intervals $T=1$
between the availability of consecutive vaccine batches with $V$ doses each.
A fraction $v(t)=V/N(t)$ from the simplex embracing $N(t)>V$ susceptible
individuals at $t$ is selected for vaccination and then, removed. The
simplex must then be re-scaled in order to map the remaining $N(t+1)=N(t)-V$
susceptible into the interval $[0,1]$ to resume the process of vaccination
at the time $t+1$. Therefore, each iteration of the simulation comprises
three steps: a decision step; a removal step; and a re-scaling step, which
are sketched in \textbf{Figure S1}.

\bigskip

\bigskip

{\Large Decision Step}

\bigskip

\bigskip

This is the part that distinguishes the strategies to drive vaccination. To
proceed with the \textit{envy-free}, we follow the procedure detailed in
Section \ref{model} to build the simplex, label it, and use Equations (\ref%
{total utility A em I} - \ref{total utility B em II}) to determine which
side each counselor would choose to vaccinate at each labeled point. Then,
by inspection, we identify all pairs of points $(p_{L},p_{R})$ between which
an \textit{envy-free} point $p^{\ast }(t)$ must be located. We choose the
average positions between $p_{L}$ and $p_{R}$ to approximate the actual $%
p^{\ast }(t)$ at each time. The structure of the simplex with the continuous
functions $\rho ^{(\eta )}(y)$ to approach the utility densities of
counselor $\eta =A,B,$ guarantees the existence of at least one \textit{%
envy-free} point between each pair $(p_{L},p_{R})$.

To build up such a function we suppose that a value $\psi _{k}^{(\eta )}$ is
attributed by counselor $\eta $ to each age group labeled $k\in \{1,...,m\}$%
, that decompose the population of the interval $[0,1]$ into $m$
sub-intervals, each of these enclosed between initial and final points,
respectively $y_{k}^{I}$ and $y_{k}^{F}$, for all $k.$ Continuity at the
frontiers between neighboring sub-intervals is assured by means of Sigmoid
functions with an additional parameter $B$ coinciding with the slope at the
origin:

\begin{equation}
G(y)=\frac{1}{1+e^{-By}}.  \tag{S1}  \label{Gy}
\end{equation}%
This allows a construction of the functions $\rho ^{(\eta )}(y)$ as%
\begin{equation}
\rho ^{(\eta )}(y)=\psi _{1}^{(\eta )}\left[ 1-G(y-y_{1}^{F})\right]
+\sum_{k=1}^{m-1}\psi _{k}^{(\eta )}\left[ G(y-y_{k}^{I})-G(y-y_{k}^{F})%
\right] +\psi _{m}^{(\eta )}\left[ G(y-y_{m}^{I})\right]  \tag{S2}
\label{ro (eta)}
\end{equation}

The normalized utility density functions $u^{(\eta )}(y)$ defined by
equation (\ref{normalized ui}) for each $\eta $ are evaluated for $\rho
^{(\eta )}(y)$ defined above.{}

A remark is in order here regarding an eventual identification of several 
\textit{envy-free} points in the simplex, at each time $t$. When this is the
case, we proceed by choosing the one leading to the smallest difference
between the benefits envisaged by the two counselors. In case of a tie, we
choose the \textit{envy-free} point leading to the greatest benefit
resulting from adding the two contributions. If the simplex would still
present more than one \textit{envy-free} point, we select one of them
randomly.

\bigskip

{\Large Removal Step}

\bigskip

The choice of an \textit{envy-free} point at the Decision Step prescribes a
set $\Lambda $ of $L(t)$ intervals $\Lambda =\{\lambda _{1}(t),...,\lambda
_{L}(t)\}$ $\ \lambda _{l}(t)\in \lbrack 0,1],$ each one of them selected
either by $C_{A}$ or by $C_{B}$ to maximize each one benefit, accounting for
the coupled contribution from the other\'{}s choice. The union of all $%
\lambda _{l}(t)$, $l\in \{1,...,L\}$, corresponds to the fraction $%
v(t)=V/N(t)$ of the population vaccinated at time $t$ which is then removed
from the interval $[0,1].$ At the end of this removal process occurring at
the interaction time $t$, the simplex turns out into a set of $L(t)+1$
disjoint intervals $s_{j}(t)=[s_{j}^{I}(t),s_{j}^{F}(t)]$, $\ j\in
\{1,..,L(t)+1\}$ which union

\begin{equation}
S(t)=\bigcup\limits_{j}s_{j}(t)\subseteq \lbrack 0,1]  \tag{S3}  \label{S(t)}
\end{equation}%
shall be re-scaled to define the simplex at the time $t+1$. The construction
is sketched in \textbf{Figure S1} for $L=2$.

\bigskip

{\Large Re-scaling Step}

\bigskip

After removing the individuals vaccinated at time $t,$ each interval $%
s_{j}(t)$ is re-scaled into a new interval referred to as $\zeta _{j}(t+1)$.
The union of all $\zeta _{j}(t+1)$ defines the new simplex $[0,1]$ over
which the former steps shall be repeated at time $t+1$:

\begin{equation}
Z(t+1)=\bigcup\limits_{j}\zeta _{j}(t+1)=[0,1]  \tag{S4}  \label{Z(t+1)}
\end{equation}

The intervals $\zeta _{j}(t+1)$ are set through the scale factor%
\begin{equation}
r(t;t+1)=\frac{N(t)}{N(t+1)}=\frac{1}{(1-v(t))}  \tag{S5}
\label{scale factor}
\end{equation}%
so that the first interval $\zeta _{1}(t+1)$ has its endpoints calculated as:

\begin{align}
\zeta _{1}^{I}(t+1)& =0  \tag{S6}  \label{zeta 1} \\
\zeta _{1}^{F}(t+1)& =\Delta s_{1}(t)r(t;t+1)  \notag
\end{align}%
where $\Delta s_{j}(t)=s_{j}^{F}(t)-s_{j}^{I}(t)$ is the size of the
interval $s_{j}(t)$. The remaining intervals $[\zeta _{j}^{I}(t+1)$, $\zeta
_{j}^{F}(t+1)]$ for all $j>1$ are set as:

\begin{align}
\zeta _{j}^{I}(t+1)& =\zeta _{j-1}^{F}(t+1)  \tag{S7}  \label{zeta j} \\
\zeta _{j}^{F}(t+1)& =\zeta _{j}^{I}(t+1)+\Delta s_{j}(t)r(t;t+1)  \notag
\end{align}

$Z(t+1)$ defines the simplex that will be considered in the next iteration,
at the time $t+1$. The three steps described above are iterated up to the
vaccination is completed.

\bigskip

{\LARGE About the choice of parameter }$d$

\bigskip

Sperner's Lemma guarantees that the envy-free strategy will always find at
least one pair of points $(p_{L},p_{R})$ enclosing an envy-free point $%
p^{\ast }$ at the end of each iteration time. However, if the considered
number $d$ of divisions of the simplex is too small, the average between
these two points may not be a good approximation to $p^{\ast }$, as we have
assumed. In this case, changing $d$ may lead to oscillations in the value of 
$p^{\ast }$ and, most probably, in the quantities derived from it. Improving
the approximation by increasing $d$ is expected to reduce such oscillations
as $p^{\ast }$ approaches its actual value. This, however, implies adding
considerable computational costs to the numerical procedure. To achieve a
compromise between mathematical accuracy and computational performance in
this case, we examine how the change in $d$ directly affects the temporal
averages of the quantities shown in \textbf{figures 4, 6, and 7} using as an
example, the population age distribution from the U.S.. \textbf{Figure S2(a) 
}shows the average temporal behavior of the difference of the benefits
(absolute values) ((\ref{time_average _difference})) due to the
contributions of the two counselors obtained through the envy-free strategy,
as $d$ increases. \textbf{Figure S2(b)} shows the corresponding behavior of
the cumulative differences (\ref{time average diff cum}), and \textbf{Figure
S2(c)} shows the mean cumulative benefit (\ref{time_average_cum}). As
expected, the amounts oscillate around the mean until stabilizing at a
certain value of $d$ that is not the same for the different quantities
analyzed. We proceed into the whole numerical calculation presented above
choosing $d=100$ which seems suitable to ensure convergence of the results
in all cases.

\bigskip

{\LARGE Other strategies}

\bigskip

The other strategies considered to simulate the dynamics of vaccination, in
particular the \textit{maximize benefit} and the \textit{oldest-first},
introduce changes into the Decision Step described above.

The \textit{maximize-benefit}, is based on the choice of the region $\Omega
(t)_{\max }$ in the simplex which is of the size of the total fraction $%
v(t)=V/N(t)$ of individuals to be vaccinated at each time $t,$ such that it
maximizes the total benefit, accounting for both counselors according to the
prescription in (\ref{U(t)}) for $\Omega (t)=\Omega (t)_{\text{max}}$. The
iterating procedure follows then the same removal and re-scaling steps as
for the \textit{envy-free}.

The implementation of the \textit{oldest-first} strategy consists in
allocating the total fraction $v(t)=V/N(t)$ of vaccine doses available at
the time $t$ to the oldest fraction of the population present at that time.
The resulting benefit is evaluated according to (\ref{U(t)}) for $\Omega
(t)=\Omega (t)_{\text{oldest}}$. The iterating procedure follows then the
same removal and re-scaling steps as for the \textit{envy-free}.

\bigskip

{\LARGE Acknowledgements}

The authors acknowledge D.H.U. Marchetti for comments and suggestions on the
manuscript. P.R.A acknowledges the financial support from the Brazilian
agency Coordena\c{c}\~{a}o de Aperfei\c{c}oamento de Pessoal de N\'{\i}vel
Superior (CAPES).

\section*{\protect\bigskip Legends}

\begin{itemize}
\item \textbf{Figure 1} - A labeled simplex defined by a partition with $d=9$%
.\ The scheme follows Ref. \cite{Playful Introduction} to illustrate in the
present case possible choices of counselors $C_{A}$, at points $A$\ and $%
C_{B,}$\ at points $B,$\ about vaccinating on side $I$\ or on side $II$.
Each of the regions between an identified pair $(p_{L},p_{R})$ encloses a
point $p^{\ast }$ that sets an envy-free division of the simplex.

\item \textbf{Figure 2} - Illustrative example. Utility density functions
evaluated by each counselor $C_{A}$\emph{\ }(purple) and $C_{B}$ (green)
considering the four combinations of utilities, as indicated, for the
population age-groups of the U.S.

\item \textbf{Figure 3 }- Simulated time series for the benefits (\ref{Ui})
acquired by each counselor, $C_{A}$\emph{\ }(orange) and $C_{B}$ (blue)
through (a) a random procedure and strategies (b) \textit{maximize-benefit},
(c) \textit{oldest-first}, (d) \textit{envy-free}. The utility density
functions employed correspond to the combination defined as Default in
Fig.(2) and the results shown are for the population age-distribution of the
U.S.

\item \textbf{Figure 4 -} Time average of the differences (absolute values) (%
\ref{time_average _difference}) between the contributions of the two
counselors to the benefits obtained through each strategy, extended to the
population age-groups of the selected countries, as specified. The utility
density functions employed correspond to the combination defined as Default.

\item \textbf{Figure 5 -} Simulated time series for the cumulative benefits $%
\Phi _{\eta }(t)$ (\ref{cumulative benefit}) acquired by each counselor, $%
C_{A}$\emph{\ }(orange) and $C_{B}$ (blue) through (a) a random procedure
and strategies (b) \textit{maximize-benefit}, (c) \textit{oldest-first}, (d) 
\textit{envy-free}. The utility density functions employed correspond to the
combination defined as Default in Fig.(2). The results shown are for the
population age-distribution of the U.S.

\item \textbf{Figure 6} - Time average of the differences (absolute values) $%
\overline{\text{ }\Delta \Phi }$\emph{\ }(\ref{time average diff cum})
between the contributions of the two counselors to the cumulative benefits
obtained through each strategy, extended to the population age-distributions
of the selected countries, as specified. The utility density functions
employed correspond to the combination defined as Default.

\item \textbf{Figure 7 -} Time average of the mean cumulative benefit $%
\overline{\Phi }$ (\ref{time_average_cum}) between the contributions of the
two counselors, for each strategy and selected countries. The utility
density functions employed correspond to the combination defined as Default.

\item \textbf{Figure 8 -} Time averages of \ (a) the cumulative differences
(absolute values) $\overline{\text{ }\Delta \Phi }$ (\ref{time average diff
cum}), and (b) the instantaneous differences (absolute values) $\overline{%
\Delta \mathcal{U}}$ (\ref{time_average _difference}) against the mean\emph{%
\ }$\overline{\Phi }$ (\ref{time_average_cum}), evaluated for the population
age-distributions of $236$ selected countries for strategies and utility
combinations, as indicated by colors. The distributions of the points are
indicated by the lateral diagrams.\textit{\ Maximize-benefit} (blue) \textit{%
minimize-benefit} (green), \textit{oldest-first} (brown), \textit{envy-free}
(red) and random procedure (orange).

\item \textbf{Figure 9 -} Indication of the population fraction per age
group to receive the doses at each time to follow the \textit{envy-free}
strategy considering the different combinations of utilities, as indicated.
The results shown are for the population age-distribution of the U.S.

\item \textbf{Figure S1 -} Schematic view of a model simplex at an iteration
time $t$. (1) An envy-free division point $p^{\ast }$\ is identified. (2)
The regions $\lambda _{1}$\ and $\lambda _{2}$\ corresponding to the
fractions of individuals that received the doses are removed from the
simplex. (3) The remaining regions $s_{1}$, $s_{2}$, and $s_{3}$\ are reset
through the scale factor $r$ to recompose the simplex for the analysis at
time $t+1$.

\item \textbf{Figure S2} - Study of convergence of the results as $d$\
varies. The quantities examined are indicated on the axis. In each case, the
median is indicated by the orange line.
\end{itemize}

\newpage

\bigskip\
\bigskip\
\bigskip\
\bigskip\

\renewcommand{\thefigure}{1}
\begin{figure}[th]
\centering
\includegraphics[scale=1.05]{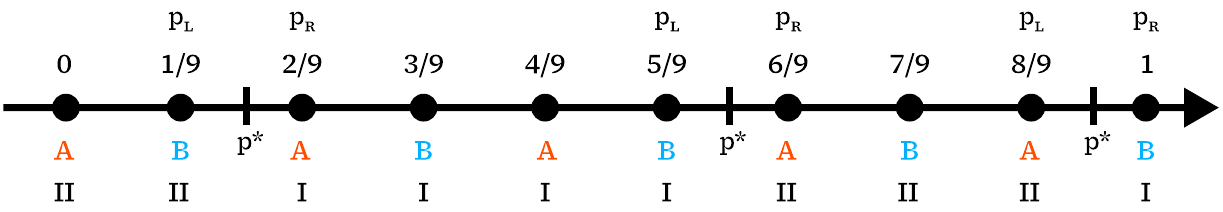} \centering
\caption{}
\end{figure}

\bigskip\
\bigskip\
\bigskip\
\bigskip\

\renewcommand{\thefigure}{2}
\begin{figure}[th]
\centering
\includegraphics[scale=0.95]{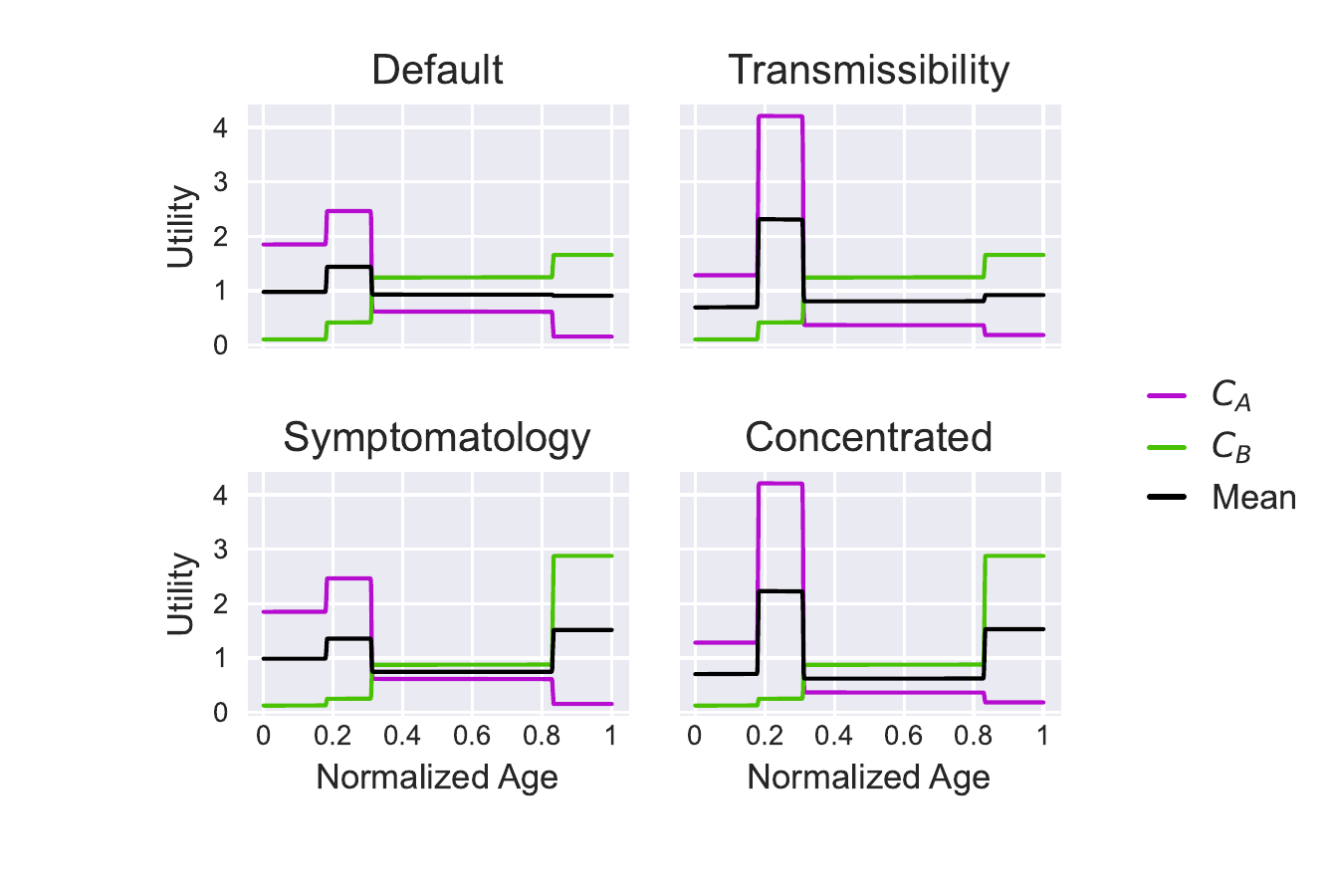} \centering
\caption{}
\end{figure}

\newpage

\bigskip\
\bigskip\
\bigskip\
\bigskip\

\renewcommand{\thefigure}{3}
\begin{figure}[th]
\centering
\includegraphics[scale=0.95]{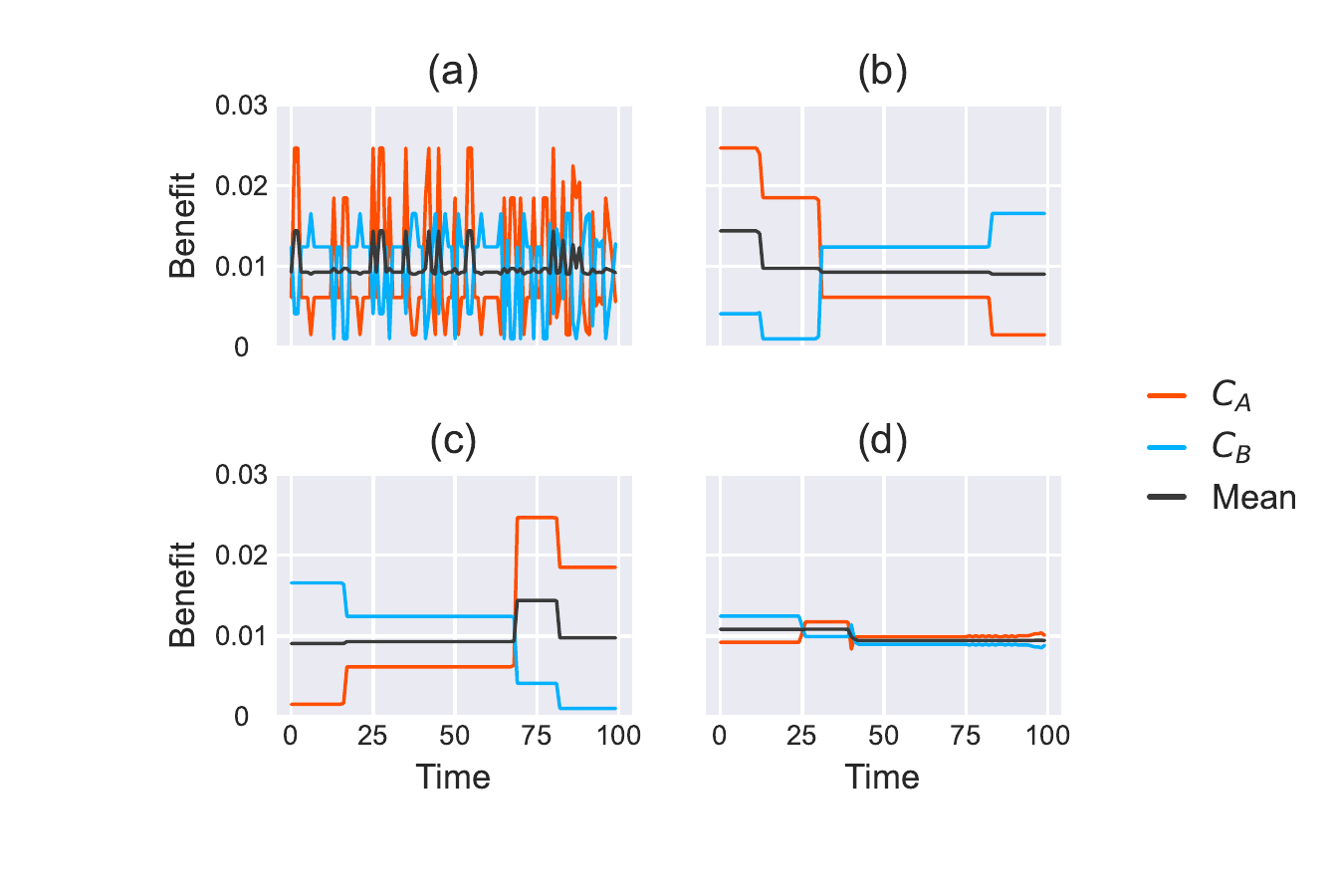} \centering
\caption{}
\end{figure}
\renewcommand{\thefigure}{4}
\begin{figure}[th]
\centering
\includegraphics[scale=0.95]{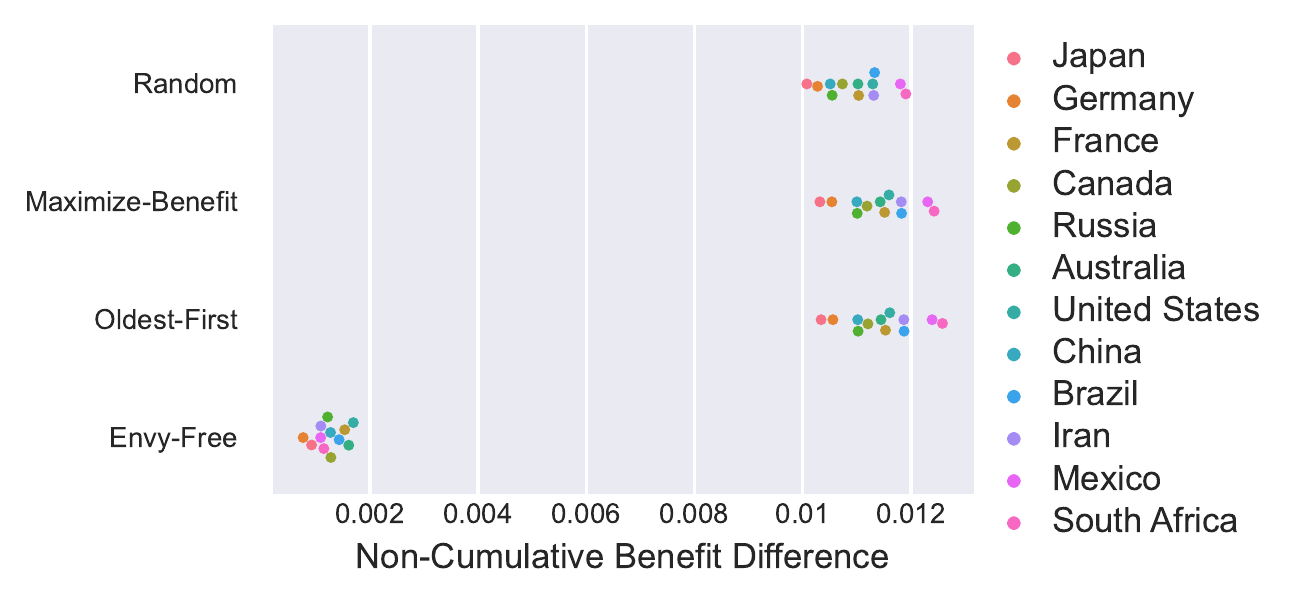} \centering
\caption{}
\end{figure}

\newpage

\bigskip\
\bigskip\
\bigskip\
\bigskip\

\renewcommand{\thefigure}{5}
\begin{figure}[th]
\centering
\includegraphics[scale=0.95]{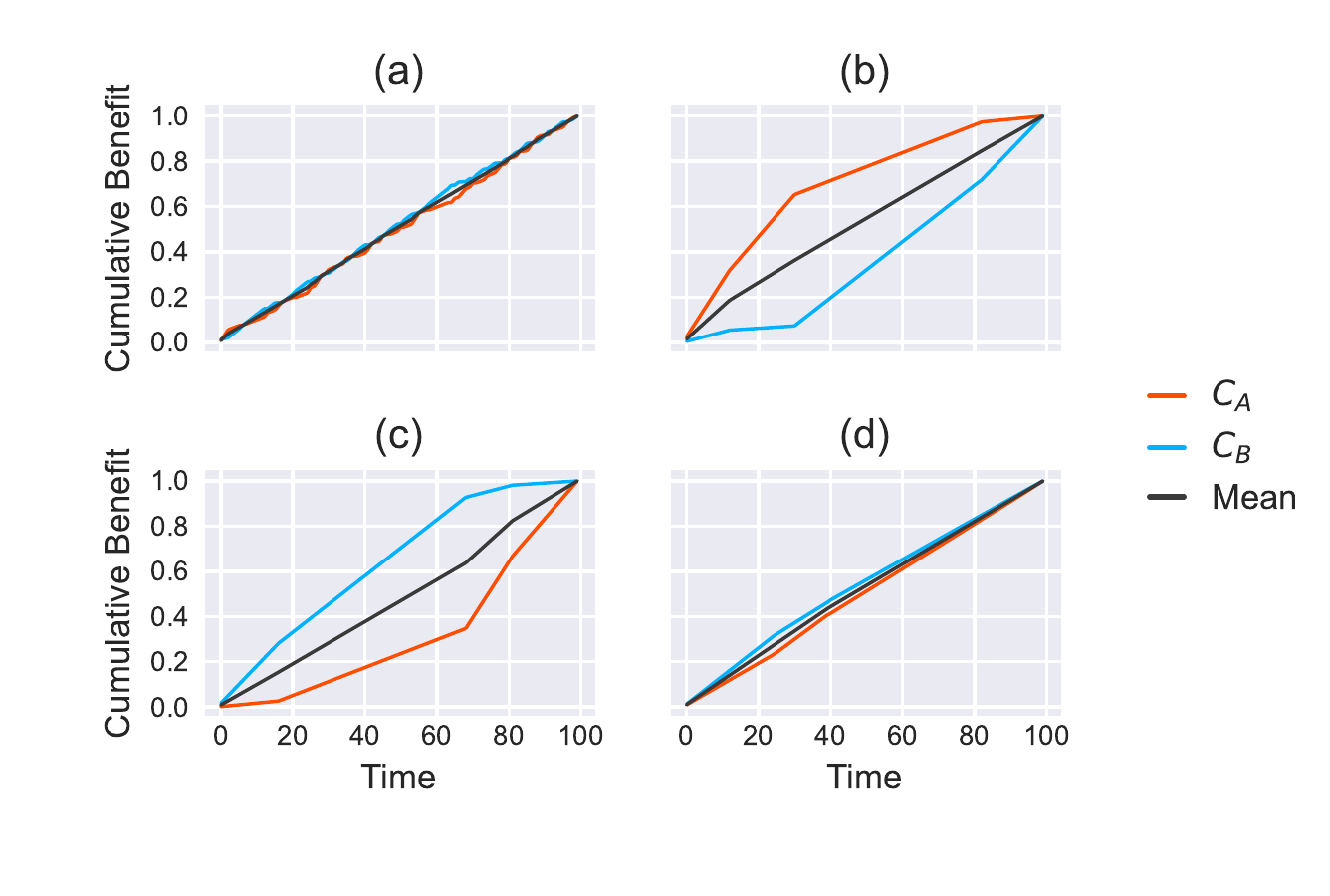} \centering
\caption{}
\end{figure}
\renewcommand{\thefigure}{6}
\begin{figure}[th]
\centering
\includegraphics[scale=0.95]{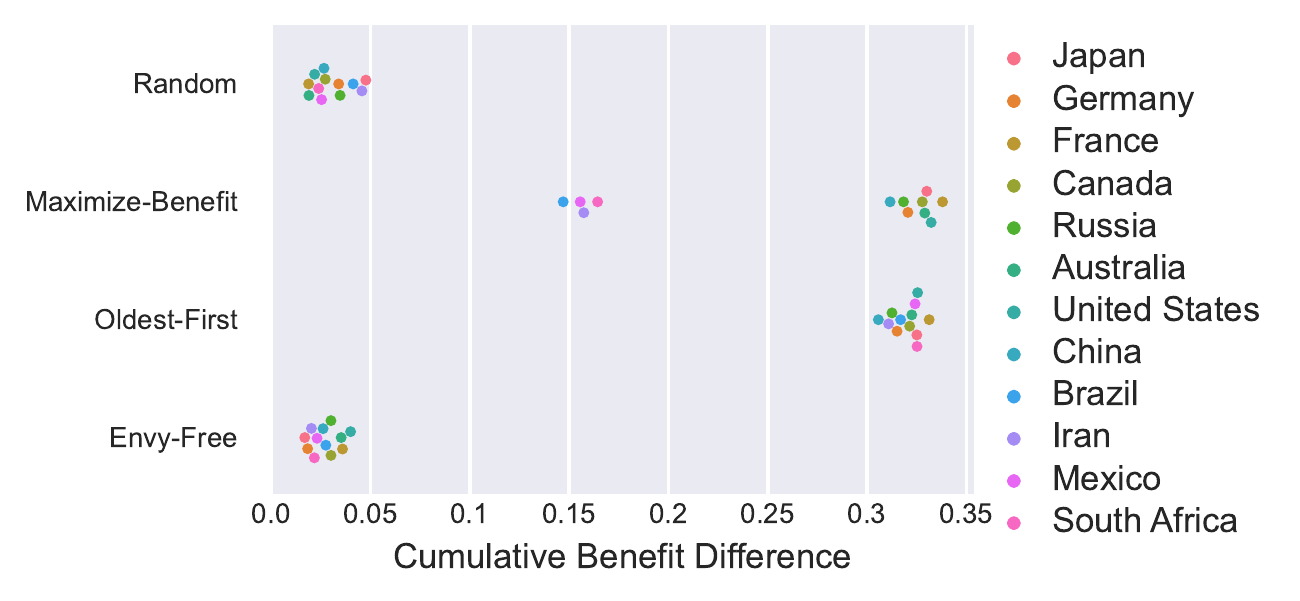} \centering
\caption{}
\end{figure}

\newpage

\bigskip\
\bigskip\
\bigskip\
\bigskip\

\renewcommand{\thefigure}{7}
\begin{figure}[th]
\centering
\includegraphics[scale=0.95]{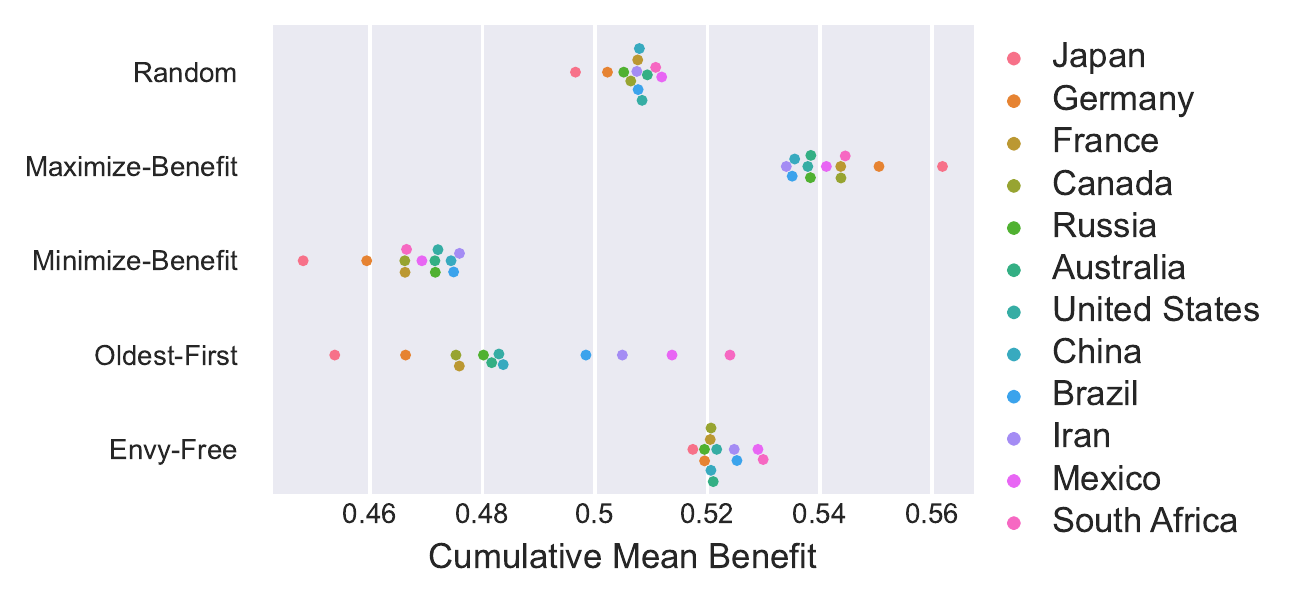} \centering
\caption{}
\end{figure}
\renewcommand{\thefigure}{8}
\begin{figure}[th]
\centering
\includegraphics[scale=0.75]{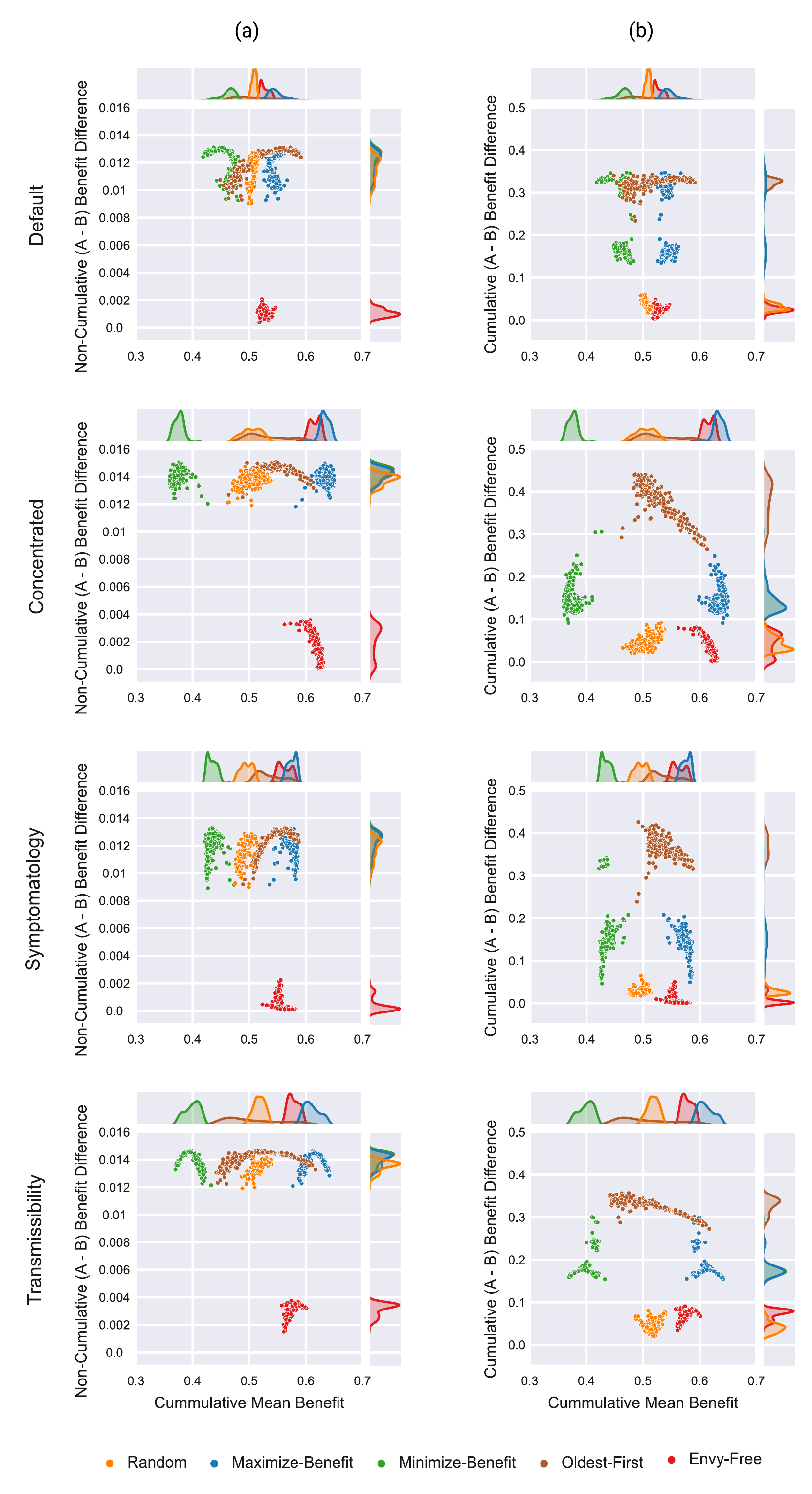} \centering
\caption{}
\end{figure}

\newpage

\renewcommand{\thefigure}{9}
\begin{figure}[th]
\centering
\includegraphics[scale=0.95]{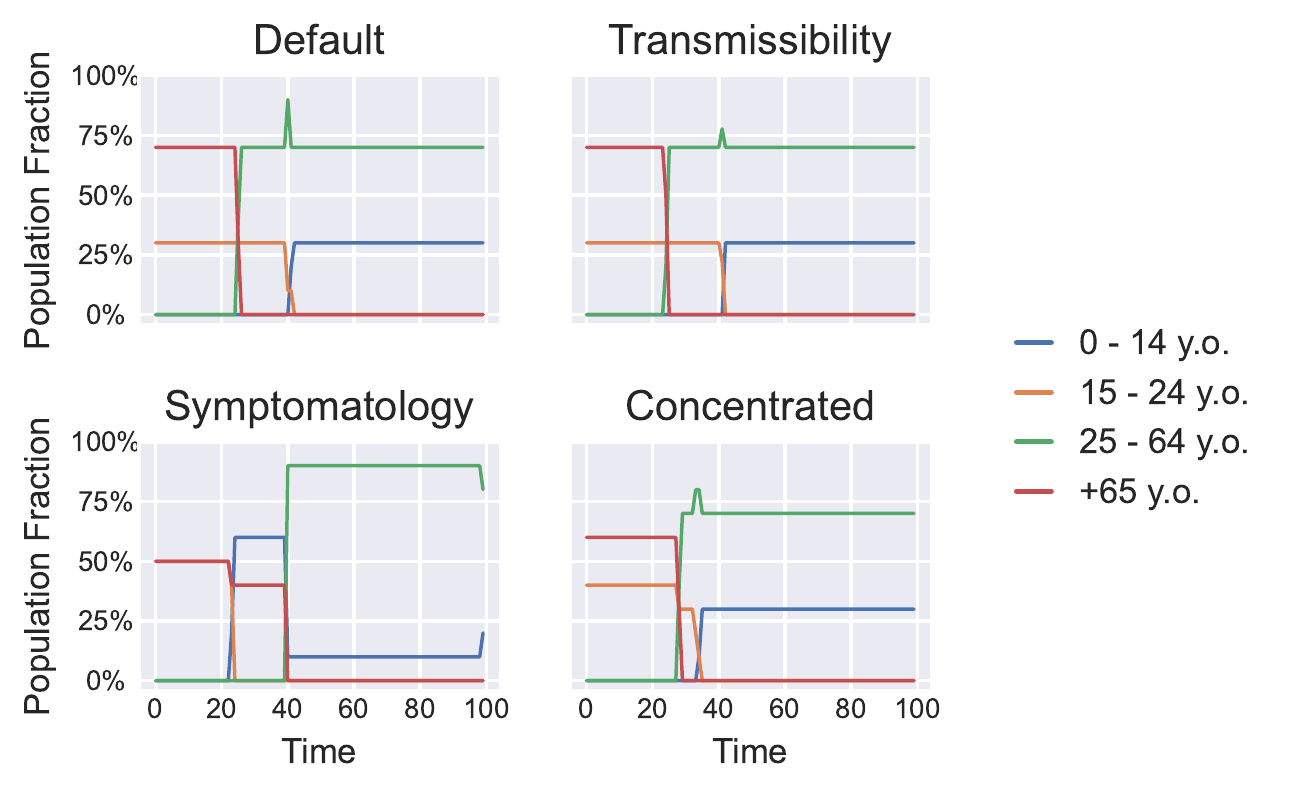} \centering
\caption{}
\end{figure}

\newpage

\renewcommand{\thefigure}{S1}
\begin{figure}[th]
\centering
\includegraphics[scale=0.45]{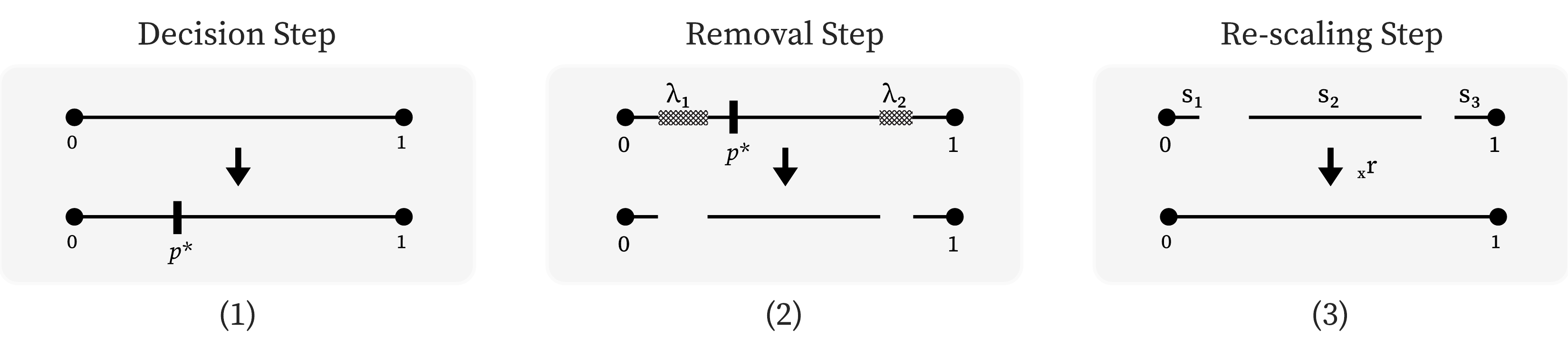} \centering
\caption{}
\end{figure}
\renewcommand{\thefigure}{S2}
\begin{figure}[th]
\centering
\includegraphics[scale=0.75]{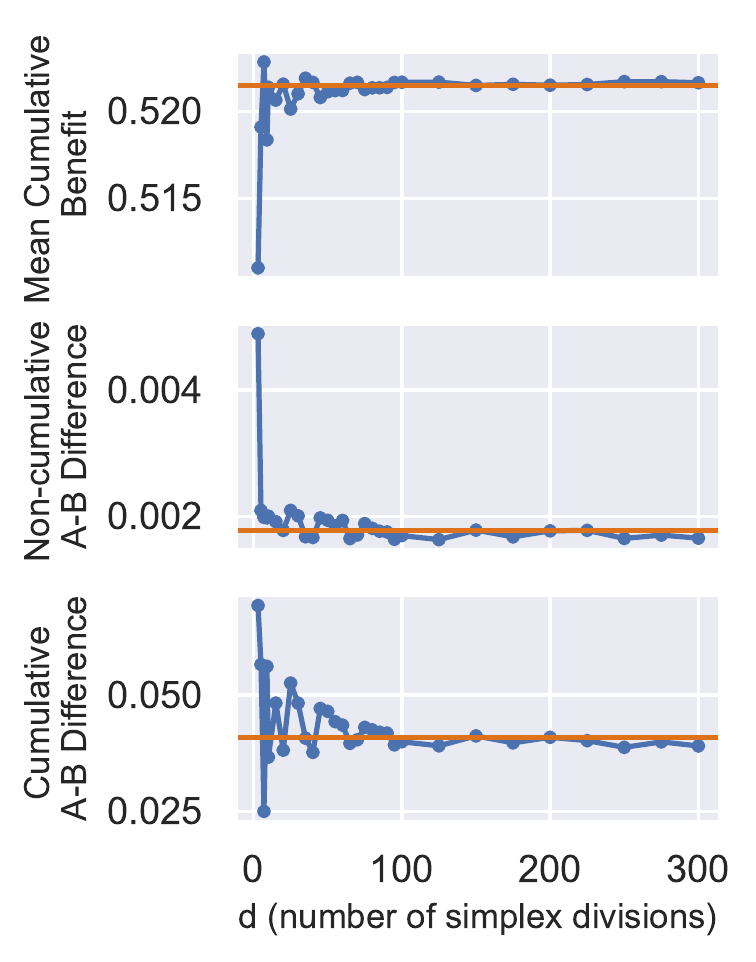} \centering
\caption{}
\end{figure}

\end{document}